\documentclass[10pt,preprint]{aastex}
\def\spose#1{\hbox to 0pt{#1\hss}}
\newcommand\lsim{\mathrel{\spose{\lower 3pt\hbox{$\mathchar"218$}}
     \raise 2.0pt\hbox{$\mathchar"13C$}}}	
\newcommand\gsim{\mathrel{\spose{\lower 3pt\hbox{$\mathchar"218$}}
     \raise 2.0pt\hbox{$\mathchar"13E$}}}

\begin{document}

\title{Broadband Observations of the Afterglow of GRB~000926: Observing the Effect
of Inverse Compton Scattering}
\author{F. A. Harrison\altaffilmark{1},
S.~A.~Yost\altaffilmark{1},
R.~Sari\altaffilmark{2},
E.~Berger\altaffilmark{1},
T.~J.~Galama\altaffilmark{1},
J. Holtzmann\altaffilmark{3},
T.~Axelrod\altaffilmark{5},
J.~S.~Bloom\altaffilmark{1},
R.~Chevalier\altaffilmark{6}, 
E.~Costa\altaffilmark{7},
A.~Diercks\altaffilmark{1},
S.~G.~Djorgovski\altaffilmark{1},
D.~A.~Frail\altaffilmark{4},
F.~Frontera\altaffilmark{7},
K.~Hurley\altaffilmark{8},
S.~R.~Kulkarni\altaffilmark{1},
P.~McCarthy\altaffilmark{9},
L.~Piro\altaffilmark{7},
G.~G.~Pooley\altaffilmark{10},
P.~A.~Price\altaffilmark{1},
D.~Reichart\altaffilmark{1},
G.~R.~Ricker\altaffilmark{11},
D. Shepherd\altaffilmark{4},
B.~Schmidt\altaffilmark{5},
F.~Walter\altaffilmark{1},
C.~Wheeler\altaffilmark{12}
}

\begin{abstract}
GRB~000926 has one of the best-studied afterglows to-date, with
multiple X-ray observations, as well as extensive multi-frequency
optical and radio coverage. Broadband afterglow observations, spanning
from X-ray to radio frequencies, provide a probe of the density
structure of the circumburst medium, as well as of the ejecta
energetics, geometry, and the physical parameters of the relativistic
blastwave resulting from the explosion. We present an analysis of {\em
Chandra X-ray Observatory} observations of this event, along with {\em
Hubble Space Telescope} and radio monitoring. We combine these data
with ground-based optical and IR observations and fit the synthesized
afterglow lightcurve using models where collimated ejecta expand into
a surrounding medium. We find that we can explain the broadband
lightcurve with reasonable physical parameters if the cooling is
dominated by inverse Compton scattering.  For this model, an excess
due to inverse Compton scattering appears above the best-fit
synchrotron spectrum in the X-ray band.  No previous bursts have
exhibited this component, and its observation would imply that the GRB
exploded in a moderately dense ($n
\sim 30$ cm$^{-3}$) medium, consistent with a diffuse interstellar cloud
environment.
\end{abstract}

\slugcomment{To appear in The Astrophysical Journal}

\altaffiltext{1}{Division of Physics, Mathematics and Astronomy, 105-24, 
California Institute of Technology, Pasadena, CA, 91125.} 
\altaffiltext{2}{Theoretical Astrophysics 130-33, California Institute of
Technology, Pasadena, CA, 91125.}
\altaffiltext{3}{Department of Astronomy, New Mexico State University, Box
30001, Department 4500, Las Cruces, NM, 88003-8001.}
\altaffiltext{4}{National Radio Astronomy Observatory, P.O. Box O, Socorro,
NM, 87801.} 
\altaffiltext{5}{Research School of Astronomy \& Astrophysics, Mount
Stromlo Observatory, Cotter Road, Weston, ACT, 2611, Australia.}
\altaffiltext{6}{Department of Astronomy, University of Virginia, 
P.O. Box 3818, Charlottesville, VA 22903}
\altaffiltext{7}{Istituto Astrofisica Spaziale, 
Consiglio Nazionale delle Ricerche, Via Fosso del Cavaliere 100, 00133 Rome, Italy.}
\altaffiltext{8}{University of California Space Sciences Laboratory, Berkeley, CA, 94720.}
\altaffiltext{9}{Infrared Processing and Analysis Center 100-22, California
Institute of Technology, Pasadena, CA, 91125.}
\altaffiltext{10}{Mullard Radio Astronomy Observatory, Cavendish
Laboratory, Madingley Road, Cambridge CB3 0HE, England UK}
\altaffiltext{11}{Center for Space Research, MIT, Cambridge, MA.}
\altaffiltext{12}{Departmant of Astronomy, University of Texas, Austin, Texas, USA}

\section{Introduction}

Broadband observations of gamma-ray burst afterglows can in principle be
used to constrain fundamental physical parameters of the explosion. In the
fireball model, a relativistic blast wave expands into the surrounding
medium, its hydrodymanical evolution being strongly influenced by the
density structure of the medium as well as by the energy content and
geometry (in particular collimation) of the ejecta. The temporal behavior of
the afterglow emission which arises from the shocked gas depends on the
shock evolution, and the partition of energy between the magnetic field and
relativistic electrons, and can therefore probe these physical parameters
given data of sufficient quality.

In this paper, we report the synthesized results from our
multi-frequency followup campaign on the relatively bright
GRB~000926. This campaign was aimed at studying the evolution of the
afterglow to constrain the model parameters described
above. \citet{phg+01}  have reported the
results from our multi-band ($BVRI$) optical monitoring. We combine
these data with 4 epochs taken with the {\em Hubble Space Telescope} WFPC2,
with {\em Chandra X-ray Observatory} (CXO) target of opportunity (TOO)
observations, and with multi-frequency radio monitoring from the Very
Large Array (VLA)\footnote{The NRAO is a facility of the National
Science Foundation operated under cooperative agreement by Associated
Universities, Inc. NRAO operated the VLA}, the Ryle Telescope, and the
Owens Valley Radio Observatory. We interpret the resulting broadband
lightcurve in the context of a theoretical afterglow model.

\section{Observations and Data Reduction}

\begin{deluxetable}{lclc}
\footnotesize 
\tablecolumns{4} 
\tablewidth{0pc} 
\tablecaption{WFPC2 HST observations of the GRB~000926 optical
afterglow. \label{tab-data}} 
\tablehead{\colhead{Date (2000, UT)} & \colhead{Filter} &
\colhead{Exposure time (sec)} & \colhead{Magnitude}}
\startdata 
Oct 7.25        & F450 & 2x1100 (1 orbit)  & 24.98 $\pm$ 0.07 \\
Oct 7.35        & F606 & 4x1100 (2 orbits) & 24.54 $\pm$ 0.03 \\
Oct 7.49        & F814 & 4x1100 (2 orbits) & 23.89 $\pm$ 0.03 \\
Oct 16.08       & F450 & 2x1100 (1 orbit)  & 25.82 $\pm$ 0.14 \\
Oct 16.18       & F606 & 4x1100 (2 orbits) & 24.27 $\pm$ 0.03 \\
Oct 16.32       & F814 & 4x1100 (2 orbits) & 24.87 $\pm$ 0.05 \\
Oct 25.05       & F450 & 2x1100 (1 orbit)  & 25.59 $\pm$ 0.12 \\
Oct 25.21       & F606 & 4x1100 (2 orbits) & 25.45 $\pm$ 0.03 \\
Oct 25.35       & F814 & 4x1100 (2 orbits) & 24.96 $\pm$ 0.05 \\
Dec 16.02       & F606 & 6x1000 (3 orbits) & 25.58 $\pm$ 0.03 \\
Dec 16.90       & F814 & 4x1100 (2 orbits) & 25.24 $\pm$ 0.07 \\
\enddata
\end{deluxetable}

\begin{deluxetable}{cccc}
\footnotesize
\tablecolumns{3}
\tablewidth{0pc}
\tablecaption{OT Magnitudes Measured by HST}\label{tab-hst}
\tablehead{\colhead{T (Days after GRB)} & \colhead{Frequency (GHz)} & \colhead{Flux $\pm \sigma$ ($\mu$Jy)}}
\startdata
10.3570 &   $6.871 \times 10^{14}$ (B) & $0.233 \pm 0.039$ \\
19.1870  &  $6.871 \times 10^{14}$ (B)  & $ -0.011 \pm 0.037 $  \\
28.2170  &  $6.871 \times 10^{14}$ (B)  & $0.057 \pm 0.039 $ \\
10.3570   &  $5.499 \times 10^{14}$ (V) & $0.257 \pm 0.023$ \\
19.1870   &  $5.499 \times 10^{14}$ (V) & $0.056 \pm 0.019 $  \\
28.2170   &  $5.499 \times 10^{14}$ (V) & $0.008 \pm 0.018$  \\
80.4670  &   $5.499 \times 10^{14}$ (V) & $-0.004 \pm 0.018 $ \\
10.3570   &  $4.673 \times 10^{14}$ (R) & $ 0.398 \pm 0.029$ \\
19.1870   &  $4.673 \times 10^{14}$ (R)  & $0.072 \pm 0.023$ \\
28.2170   &  $4.673 \times 10^{14}$ (R)  & $0.034 \pm 0.022$ \\
80.4670  &   $4.673 \times 10^{14}$ (R) & $-0.004 \pm 0.021$ \\
10.3570   & $3.806 \times 10^{14}$ (I)  & $0.525 \pm 0.058$ \\
19.1870   & $3.806 \times 10^{14}$ (I) & $0.075 \pm 0.044$ \\
28.2170   & $3.806 \times 10^{14}$ (I) & $0.053 \pm 0.042$ \\
80.4670  &  $3.806 \times 10^{14}$ (I) & $-0.012 \pm  0.048$ \\
\enddata
\end{deluxetable}

\begin{deluxetable}{cccc}
\footnotesize
\tablecolumns{3}
\tablewidth{0pc}
\tablecaption{X-ray Transient Flux}\label{tab-xray2}
\tablehead{\colhead{T (Days after GRB)} & \colhead{Frequency (GHz)} &
\colhead{Flux $\pm \sigma$ ($\mu$Jy $\times 10^2$ )}}
\startdata
2.7700 & $1.14 \times 10^{17}$ & $ 3.98 \pm 0.594$ \\
13.477 & $1.14 \times 10^{17}$ &  $ 0.156 \pm 0.035$ \\
2.2780 &  $7.5 \times 10^{17}$ &  $ 1.78 \pm 0.621$ \\
2.7700 & $7.5 \times 10^{17}$  &  $ 0.684 \pm 0.070$ \\
13.477  & $7.5 \times 10^{17}$ &  $ 0.029 \pm 0.016$ \\
\enddata
\end{deluxetable}

\begin{deluxetable}{ccccc}
\footnotesize
\tablecolumns{3}
\tablewidth{0pc}
\tablecaption{X-ray Afterglow of GRB000926}\label{tab-xray}
\tablehead{\colhead{Pointing} & \colhead{Epoch (2000, UT)} & \colhead{Band (keV)} &
\colhead{Flux $ \pm \sigma $(erg/cm$^2$/s $\times 10^{13}$)}}
\startdata
SAX        & Sep. 29.03 -- 29.53 &  1.5 -- 8 & $2.23 \pm 0.77$ \\
CXO - P1   & Sep.29.674 -- 29.851 &  0.2 -- 1.5 & $0.614 \pm 0.063$ \\
CXO - P1   & Sep.29.674 -- 29.851 &  1.5 -- 8 & $0.939 \pm 0.14 $ \\
CXO - P2   & Oct.10.176 -- 10.760 &  0.2 -- 1.5 & $0.0263 \pm 0.008 $ \\
CXO - P2   & Oct.10.176 -- 10.760 &  1.5 -- 8 & $0.0364 \pm 0.019 $ \\
\enddata
\end{deluxetable}

\begin{deluxetable}{lcc}
\footnotesize
\tablecolumns{3}
\tablewidth{0pc}
\tablecaption{Radio Observations of GRB000926}\label{tab-radio}
\tablehead{\colhead{Epoch (2000, UT)} & \colhead{Frequency (GHz)} &
\colhead{Flux  $\pm \sigma$ ($\mu$Jy)}}
\startdata
Sep 29.708 &    98.48   &  $3410 \pm\ 1020$        \\
Oct 1.708 &     98.48   &  $1890 \pm\ 750$ \\ \hline
Oct 4.186 &     22.5 &  $1415 \pm\ 185$ \\
Oct 5.216 &     22.5 &  $1320 \pm\ 240$ \\
Oct 16.721 &    22.5 &  $480 \pm\ 230$  \\  \hline
Sep 28.81 &     15.0 &    $490 \pm\ 230$  \\
Sep 30.91 &     15.0 &    $-320 \pm\ 520$ \\
Oct 1.69 &      15.0 &    $820 \pm\ 390$  \\
Oct 5.751 &     15.0 &    $460 \pm\ 330$  \\
Oct 11.751 &    15.0 &    $340 \pm\ 220$  \\ \hline
Sep 28.17 &     8.46 &  $666 \pm\ 60$   \\
Sep 28.97 &     8.46 &  $150 \pm\ 55$   \\
Sep 29.726 &    8.46 &  $368 \pm\ 26$   \\
Oct 4.186 &     8.46 &  $440 \pm\ 34$   \\
Oct 5.216 &     8.46 &  $566 \pm\ 34$   \\
Oct 7.771 &     8.46 &  $564 \pm\ 76$   \\
Oct 8.291 &     8.46 &  $143 \pm\ 77$   \\
Oct 10.281 &    8.46 &  $242 \pm\ 130$  \\
Oct 12.771 &    8.46 &  $644 \pm\ 126$  \\
Oct 15.681 &    8.46 &  $379 \pm\ 36$   \\
Oct 23.161 &    8.46 &  $277 \pm\ 34$   \\
Oct 27.131 &    8.46 &  $170 \pm\ 79$   \\
Oct 30.181 &    8.46 &  $192 \pm\ 41$   \\
Nov 26.64 &     8.46 &  $143 \pm\ 35$   \\
Dec 18.95 &     8.46 &  $160 \pm\ 21$   \\
Jan 29.44 &     8.46 &  $10 \pm\ 40$    \\
Feb 9.73  &     8.46 &  $71 \pm\ 12$    \\ \hline
Sep 28.17 &     4.86 &  $90 \pm\ 67$    \\
Sep 28.97 &     4.86 &  $100 \pm\ 45$   \\
Sep 29.726 &    4.86 &  $280 \pm\ 29$   \\
Oct 4.186 &     4.86 &  $248 \pm\ 30$   \\
Oct 7.741 &     4.86 &  $395 \pm\ 61$   \\
Oct 8.701 &     4.86 &  $370 \pm\ 70$   \\
Oct 30.201 &    4.86 &  $210 \pm\ 33$   \\
Nov 18.03 &     4.86 &  $131 \pm\ 45$   \\
Jan 6.53 &      4.86 &  $62 \pm\ 42$    \\
Feb 2.47  &     4.86 &  $54 \pm\ 41$    \\
Feb 19.28 &     4.86 & $126 \pm\ 23$     \\   \hline
Dec 16.58 &     1.43 &  $96 \pm\ 46$    \\
\enddata
\end{deluxetable}

The Interplanetary Network discovered GRB~000926 on 2000 Sep 26.993 UT
\citep{hmg+00}. The afterglow of this 25~s long event was identified
less than a day later \citep{gcc+00,dfp+00}. The redshift, measured
from optical absorption features, is $2.0369 \pm 0.0007$
\citep{fmd+00,cdk+00}. The afterglow was well-monitored in the
optical \citep{phg+01,fyn+01}, and was detected in the IR
\citep{dip+00,fyn+01}. Here we describe {\em Hubble Space Telescope
(HST), Chandra X-ray Observatory (CXO)} and radio observations.

\subsection{HST Observations}

As part of an {\em HST} cycle 9 program we observed GRB~000926 at four
epochs with the Wide Field Planetary Camera 2 (WFPC2), with the OT placed on
WFPC CCD\#3. In the first three epochs we observed at three passbands
corresponding to the F450W, F606W, and F814W filters, and in the final epoch
we used only the F606W and F814W filters. These observations took place
between Oct 7.25 (10.26 days after the GRB) and Dec 16.9 (81.9 days after).

Table~1 shows a log of the HST observations, along with the magnitude
derived for each filter for the 2-pixel radius region surrounding the
OT. We determined the aperture correction using a 2 -- 5 pixel radius,
and we quote the corresponding 5 pixel radius magnitude. We calibrated
the zeropoints, and converted the WFPC2 filters to Johnson Cousins
magnitudes using the color transformations from \citet{hbc+95}. We
estimate the associated calibration uncertainty to be about 0.10
magnitude in $B,V,$ and $R$ bands, and 0.20 magnitudes in the $I$
band. To perform the transformations, we have interpolated the
3-filter WFPC2 data to obtain photometric points in $B,V,R$ and $I$.
Measurements in only three of the bands are, therefore, truly
independent. We tied the WFPC2 calibration to the ground-based data of
\citet{phg+01} using their tertiary star at position RA~$%
12^{h}04^{m}11^{s}.75$ DEC~$+51^{\circ }46^{\prime }56^{\prime \prime
}.1$ (J2000). The magnitudes we derive for this star agree with those
obtained by
\citet{phg+01} to within 0.10~mag in $B,V$, and $R$, and to
within $0.20$~mag in $I$. \citet{gal+01} provide
further details on the image reduction and magnitude calibration.

From flattening in the late-time ($t\gtrsim 10$ days) optical
lightcurve we infer that the HST data are contaminated by a contribution
from the host galaxy.  We do not actually resolve the host in our HST
images, and we observe flattening even with our small (0.14$^{\prime
\prime }$) extraction radius.  This implies that the host itself is
compact, or that the constant emission is from a bright knot or
filament associated with the host.  Future HST observations scheduled
for late May 2001 may resolve this feature, and we will present these
results elsewhere.  To determine the flux from the OT we subtract this
constant contribution.  

To derive the magnitude for
the constant term, we fit the ground-based optical lightcurve from
Price {\em et al}.  (2001) combined with our HST data to a decaying
function with a constant added: 
\begin{equation}
F(t,\nu )=F_{0}\ \nu ^{\beta }\ \left( (t/t_{\ast })^{-\alpha
_{1}s}+(t/t_{\ast })^{-\alpha _{2}s}\right) ^{-1/s} + F_{\rm host}(\nu).
\label{eq-function}
\end{equation}
The decaying function, taken from \citet{bhr+99}, has no physical
significance, but provides a simple and general parametric description
of the data. Here $\alpha _{1}$ and $\alpha _{2}$ are the early and
late time asymptotic temporal slopes respectively, $t_{\ast }$ is
the time of the temporal slope break, $s$ is a parameter that
determines the sharpness of the transition, and $\beta $ is the spectral
slope. From the combined ground-based
and HST\ data we derive $
\alpha _{1}= 1.46 \pm 0.11$,\,$\alpha_{2}= 2.38 \pm .07$,\,$t_{\ast }=1.8 \pm 0.1$ days, 
$\beta =-1.49 \pm 0.07$. All errors in this paper are 1-$\sigma$.
The best-fit value for the constant is $F_{host}(B)=25.77\pm 0.14$,
$F_{host}(V)=25.60\pm 0.1$, $ F_{host}\left( R\right) =25.38\pm 0.1$,
$F_{host}\left( I\right) =25.12\pm 0.10$. We note that our fit values
are similar to those derived by Price et al. (2001), but the
late-time, high-resolution HST data allow us to determine the flux
from the compact region near the OT.

We derive the OT magnitudes by subtracting the constant term,
$F_{host}$, given above from each band, correcting for the foreground
Galactic extinction of $E_{B-V}=0.0235$~mag
\citep{sfd98}. Table~2 contains the resulting
dereddened, host-subtracted OT flux values converted to $\mu $Jy.  The
measurement errors include a contribution (1--$\sigma $) of 0.026(B),
0.016 (V), 0.018 (R), 0.020 (I) $\mu $Jy added in quadrature to the
statistical and systematic calibration errors to reflect the
uncertainty in the constant host term derived from the fit.

\subsection{CXO Observations}

The {\em BeppoSAX} Medium Energy Concentrating Spectrometer (MECS)
instrument (sensitive from 1.6 -- 10~keV) discovered the X-ray afterglow of
GRB~000926 in an observation made on Sep 29.03 -- 29.52 \citep{pir+00}.
\citet{ggp+00} observed the source for 10~ksec as part
of a Cycle~1 CXO program using the ACIS S3 backside-illuminated chip on Sep
29.674 -- 29.851 (referred to here as CXO-P1). Supporting our HST cycle 9
program, CXO again observed GRB~000926 in a 33-ksec long TOO taken Oct
10.176--10.76 UT (CXO-P2), also with the transient placed on the ACIS S3
backside-illuminated chip. The afterglow was clearly detected in each of
these observations. The transient was not detected in the {\em BeppoSAX} Low
Energy Concentrating Spectrometer (co-aligned with the MECS, and sensitive
in the 0.1 -- 10~keV band), however the effective exposure time was only
5~ksec. \citet{pir+01} present a detailed spectral
analysis of the X-ray data, including combined fits to all pointings.

We analyzed the data from CXO-P1 and CXO-P2 using software provided by the 
{\em Chandra} X-ray Observatory Center (CXC) to filter the events, extract
source counts and subtract background. For the latter we used an annular
region surrounding the source. We fit the two {\em Chandra} pointings
separately using a powerlaw model plus absorption. The best fit photon
spectral index for the CXO-P1 data is $\alpha = 1.9 \pm 0.21$ and for CXO-P2
is $\alpha=2.23 \pm 0.34$, with best-fit hydrogen column depth $N_H = (4.8
\pm 3) \times 10^{20}$~cm$^{-2}$ (CXO-P1) and $N_H = (3.0 \pm 2.5) \times
10^{20}$~cm$^{-2}$ (CXO-P2). We find no evidence for additional
absorption in the X-ray spectrum at the redshift of the host ($z =
2.04$). This is consistent with the host $A(V)$ of 0.1~mag derived
from model fits to the optical data (see below), if we adopt typical dust to
gas ratios. For an LMC-like
extinction curve, for example, this $A(V)$ corresponds to column depth
of $N_H = 2 \times 10^{20}$~cm$^{-2}$ for a host at $z =
2.04$. Given the host redshift, such a low column is not detectable in the
X-ray spectrum, and so no correction for absorption in the GRB host is
necessary.

The detection statistics in the X-ray are limited, so for the purposes of
our modeling we divided the energy range into soft (0.2 -- 1.5 keV) and hard
(1.5 -- 8 keV) bands. We converted counts to flux using exposure maps
weighted in energy using a photon powerlaw spectral index of $\alpha = 2$
and $N_H = 2.7 \times 10^{20}$~cm$^{-2}$, corresponding to the
value for our own Galaxy, as determined by W3nH\footnote{%
W3nH is available at http://heasarc.gsfc.nasa.gov.}.  We adopt this
value since it is consistent with
that derived from our spectral modeling, and we find no significant evidence
for additional absoprtion in the host. For the {\em SAX} observation, we
used the best fit spectral model to determine the hard band flux (the MECS
response does not allow a soft-band flux to be determined).

Table~3 shows the flux values, not corrected for
Galactic absorption, for the three observations. To determine the
center energy for the band, we took the mean, weighted using a photon
index of $\alpha = 2$.  We determined flux errors by adding in
quadrature the statistical error and the error due to the uncertainty
in spectral slope. Since the flux is decaying during each observation
interval, we weight the time of the observation by $t^{-2}$ and
average to determine the mean epoch. Table~4 lists the
center frequencies and flux values for the X-ray transient (converted
to $\mu$Jy), where we have corrected for Galactic absorption.

\subsection{Radio Observations}

We obtained radio observations at frequencies from 1.43 to 100 GHz at
a number of facilities: the Owens Valley Radio Observatory
Interferometer (OVRO), the Ryle Telescope, and the VLA. Table~5
summarizes these observations, organized by frequency. The 98.48~GHz data
were taken at OVRO, the 15~GHz data at Ryle, and all other frequencies
were observed using the VLA. In performing the observations, reducing
the data and deriving flux errors we adopted the methodology described
in detail in \citet{kfs+99}, \citet{fra+00}, and
\citet{bsf+00}.  For the OVRO observations, we used the quasar 1726+455 
for phase calibration and Uranus for absolute flux calibration.

\section{Afterglow Model}

\begin{figure}
\plottwo{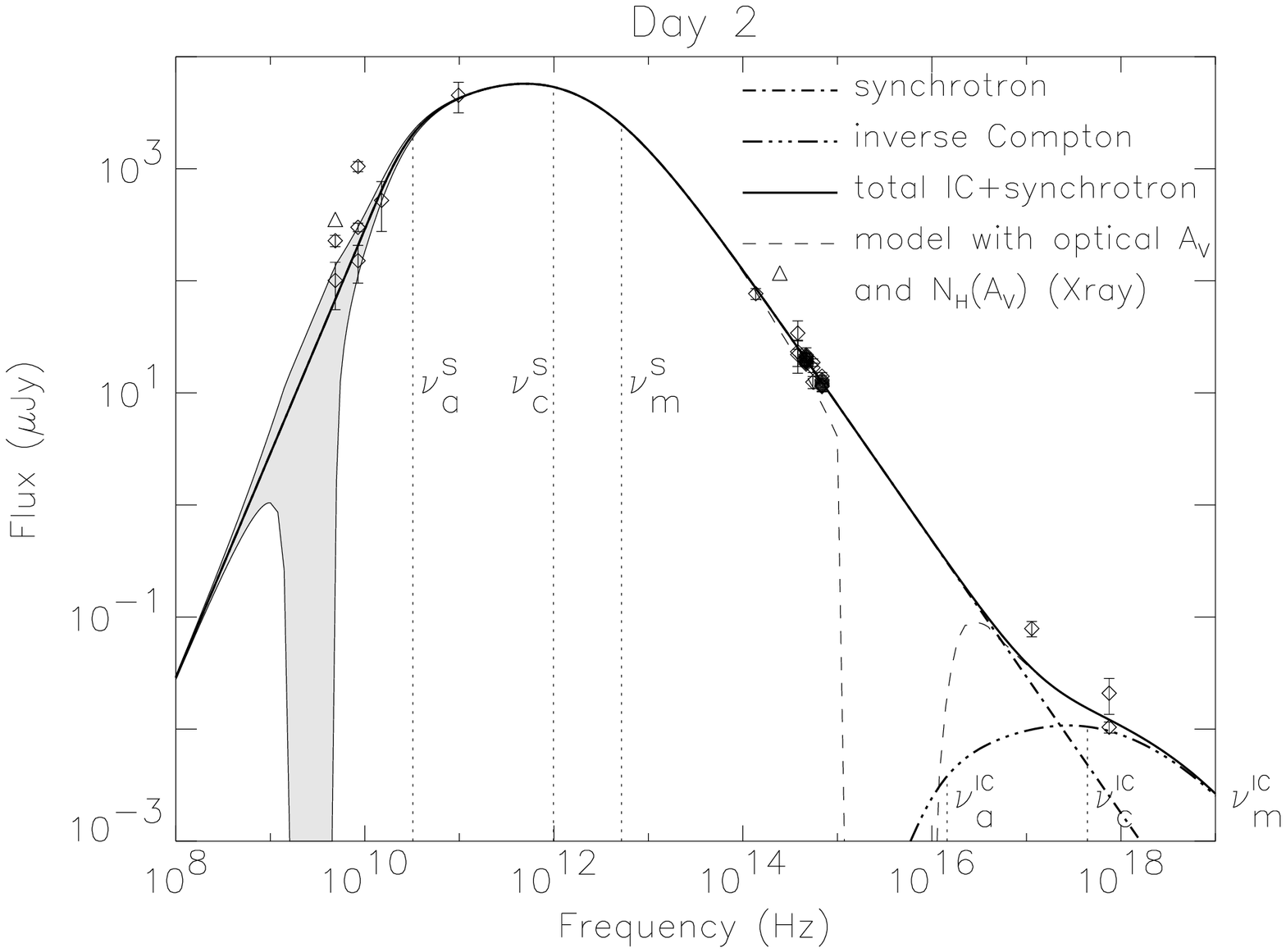}{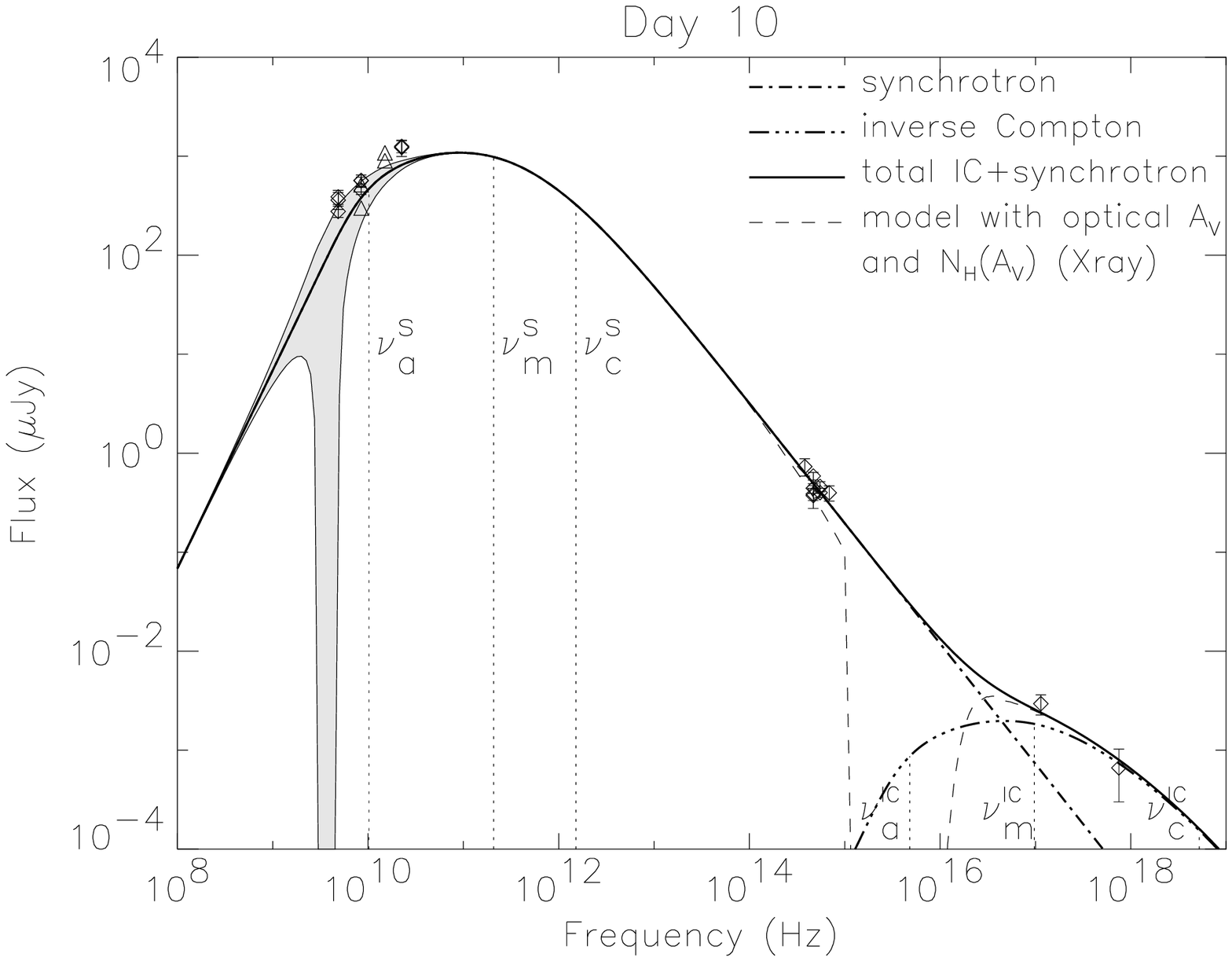}
\caption{(a) The broadband spectrum of the GRB~000926 afterglow
on day 2 (after the jet break).  All data taken between 1 -- 3 days
are included, where we have used the model calculations to extrapolate
the points forward or backward in time.  
The relative error seen in the graph therefore represents 
the true devation of the observed data from the model at the exact
time of the observations. The data, corrected for host extinction, are shown by
diamonds, with $1-\sigma$ errors, the solid line is the best-fit
model without host extinction, and the dashed line shows model plus
extinction.  The locations of the synchrotron spectral
breaks $\nu_a^{s}, \nu_c^{s}$, and $\nu_m^{s}$ associated with
self-absorption, cooling, and the maximum electon energy are
indicated.  We also indicate the corresponding breaks for the IC
spectrum (see Sari \& Esin (2001) for details).
{\bf (b)} The broadband spectrum of the GRB~000926 afterglow
on day 10 (after the jet break). \label{fig1}}
\end{figure}

\begin{figure}
\epsscale{0.75}
\plotone{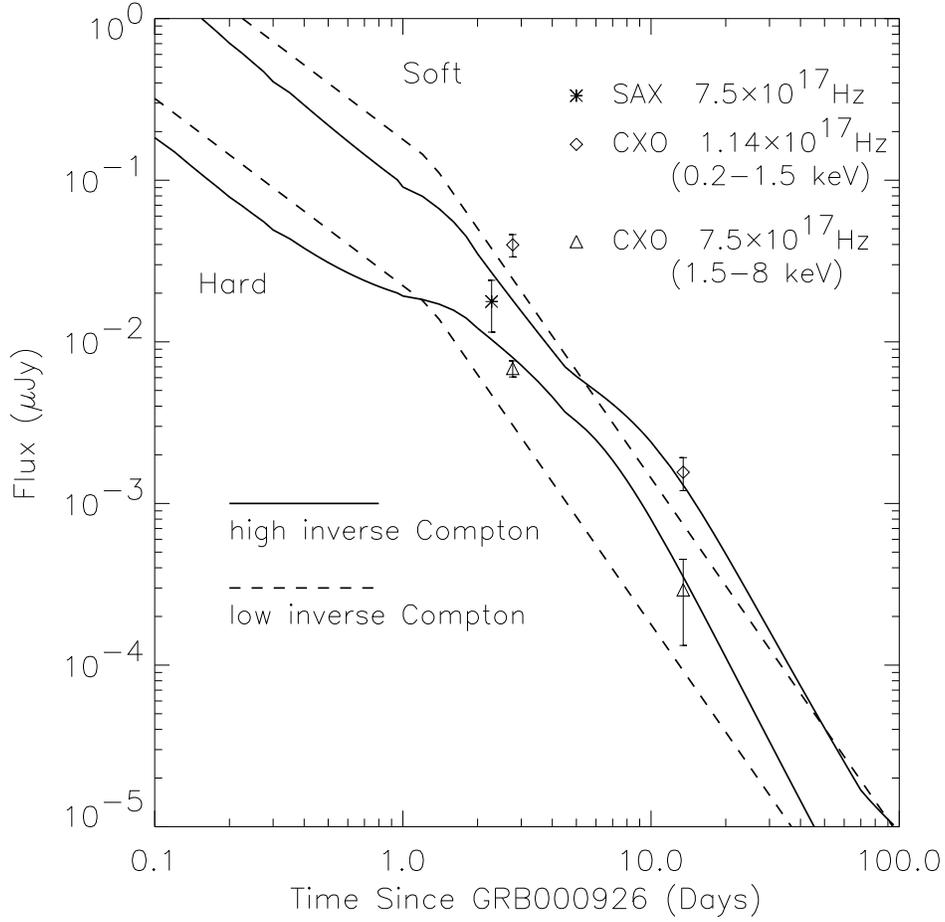}
\caption{The X-ray afterglow lightcurve from {\em BeppoSAX} and
{\em Chandra}.  The {\em Chandra} data have
been broken into two bands, hard (1.5 -- 8 keV) and soft (0.2 -- 1.5
keV), with center frequencies (weighted with a photon spectral index
of $\alpha = 2$) of $7.5 \times 10^{17}$ Hz and $1.14 \times 10^{17}$
Hz respectively.  The data have been corrected for absorption in our Galaxy
(column of $N_H = 2.65 \times 10^{20}$~cm$^{-2}$).  We also show the
model calculations for both high- (solid line) and low-IC (dashed line) constant-density ISM models.}
\end{figure}

\begin{figure}
\plotone{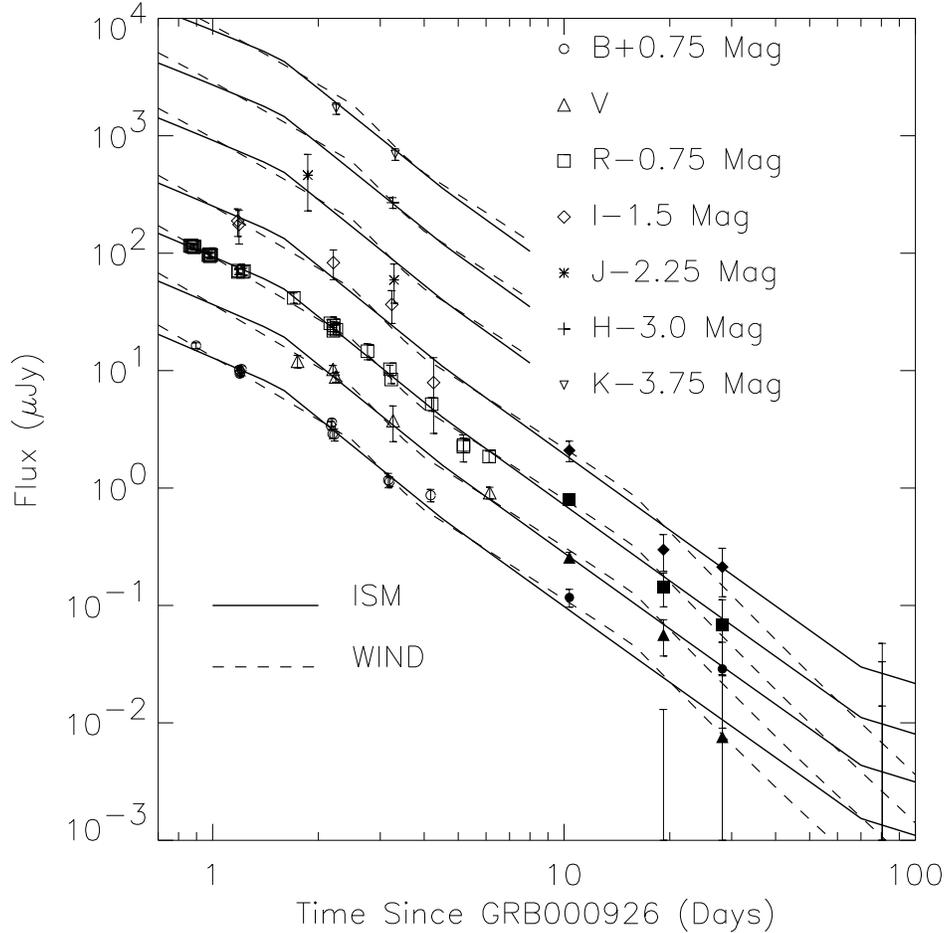}
\caption{The optical/IR afterglow lightcurve in seven bands.  We
have subtracted the host contribution, as determined from the
late-time {\em HST} data, and corrected for Galactic extinction.  The
ISM (solid lines) and best-fit wind (dashed line) models include the effect of
extinction in the host galaxy, as determined by the best-fit value
using an extinction curve (dust to gas ratio) like that of the LMC.}
\end{figure}

\begin{figure}
\plotone{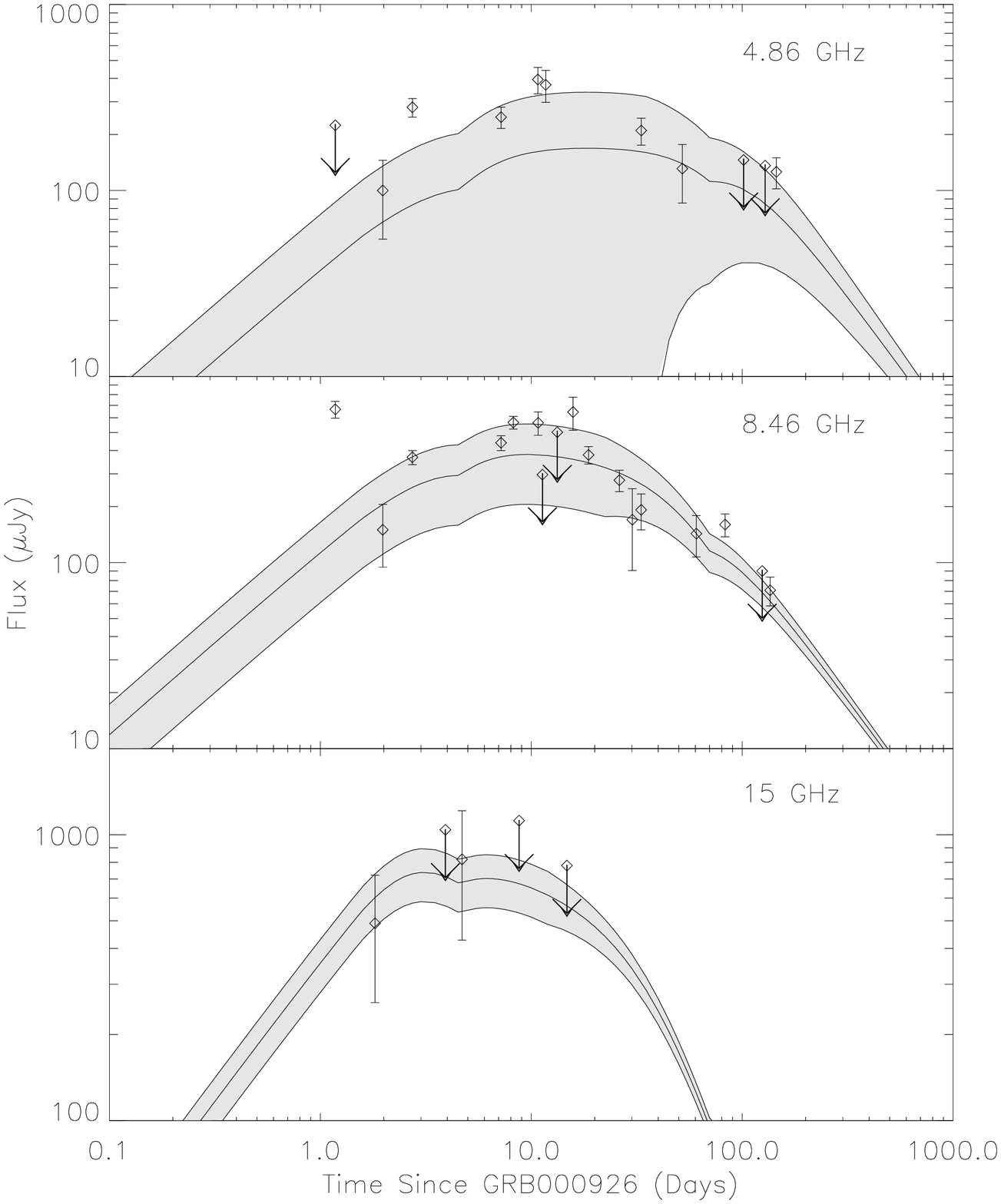}
\caption{The radio lightcurve at three frequencies.  The solid
lines show the model calculations based on the high-IC ISM
solution. The shaded regions indicate the estimated scintillation
envelopes based on the calculated size of the
fireball. The upper limits are plotted as 3-$\sigma$.}
\end{figure}

To interpret the broadband lightcurve we employ a model for the emission
from the relativistic shock (resulting from the GRB explosion) as it expands
into the surrounding medium. Our modeling is quite comprehensive. It allows
for either a constant density medium (which we refer to as ISM), or for a
density depending on radius as $\rho = A r^{-2}$ referred to as the wind
model, since this profile would be produced by a stellar progenitor's wind 
\citep{cl99}. We also allow for either isotropic or collimated ejecta, and
we fit for extinction in the host galaxy. We include the effects of inverse
Compton scattering on the evolution of the synchrotron spectrum, as well as
the contribution of IC emission to the observed spectrum. IC scattering is
not included in most models used to fit GRB afterglows, but if the ambient
medium is of sufficiently high density, it becomes an important effect.
Compton emission can dominate the total afterglow cooling rate for months
after the event, and can even be directly observed in the X-ray \citep{pk00,se01}.

We calculate the shock emission at a given time and frequency from a
number of fundamental parameters: the initial isotropic-equivalent
energy, $E$ in the shock, the electron powerlaw slope, $p$, the
electron and magnetic field energy fractions, $\epsilon_e$ and
$\epsilon_B$, and the density of the circumburst medium. We employ the
equations given by \citet{spn98} and \citet{gps99a,gps99b} to
determine the evolution of the synchrotron spectrum for the constant
density medium, and the equations given by \citet{cl99,cl00} for the
wind model. We include the effect of inverse Compton scattering using
the treatment given by \citet{se01}. We calculate the time at which
collimation becomes evident ($\theta = 1/\Gamma$, where $\Gamma$ is
the Lorentz factor of the ejecta) and the subsequent spectral
evolution using \citet{sph99}. For the host extinction we use the
parameterization of A(V) of Cardelli et.~al. (1989) and Fitzpatrick \&
Massa (1998), joined using the formula from Reichart~(1999) in the
rest-frame optical/UV, and absorption cross-sections given by
\citet{mm83} in the rest-frame X-ray.  We consider both LMC or
SMC-like extinction laws. To fit the broadband lightcurve we vary the
fundamental parameters, iterating to minimize $\chi^2$.

At any given time the synchrotron emission, which dominates at most
energies, has a spectrum characterized by a number of breaks: a
low-frequency roll over due to self-absorption at frequency $\nu_a$; a peak
at frequency $\nu_m$ due to the peak energy in the electron distribution;
and a cooling break at $\nu_c$, where radiative energy losses become a
significant fraction of the electron energy. These breaks evolve in time as
the shock evolves, producing frequency-dependent breaks in the lightcurve.
For the case of collimated ejecta, the lightcurve will begin to steepen
when $\theta$ is smaller than the angle into which the
emission is beamed due to relativistic effects ($1/\Gamma$, where $\Gamma$
is the Lorentz factor). We use a sharp transition for this ``jet break'', and
employ spherical evolution laws up until the time of the break. After the
jet transition we assume the ejecta expand laterally, and we employ the
appropriate asymptotic formulae  until the time the ejecta
become non-relativistic. The transition to the non-relativistic evolution is
again treated as sharp.

The inverse Compton component is also characterized by a series of
spectral breaks. With the synchrotron spectrum as the photon source,
Sari \& Esin (2001) find that the upscattered component is
characterized by three breaks at $\nu_a^{IC}\sim
2{\rm min}(\gamma_c,\gamma_m)^2\nu_a$, $\nu_m^{IC} \sim 2\gamma_m^2\nu_m$, $\nu_c^{IC} \sim
2\gamma_c^2\nu_c$ where $\gamma_m$ is the characteristic Lorentz
factor of the electrons emitting with peak frequency $\nu_m$. We adopt
a spectral shape for the IC component similar to the synchrotron
spectrum, but with the breaks given by the values in Sari
\& Esin (2001). This provides a good approximation to the IC component
spectral shape except for at $\nu < \nu_a^{IC}$, but for our data this 
is not a concern (see below). Between $\nu_m^{IC}$ and $\nu_c^{IC}$ there is
only a logarithmic correction, which we ignore. 

\section{Fits to the Broadband GRB~000926 Lightcurve}

We fit the broadband GRB~000926 lightcurve employing the afterglow
model described above. In addition to the data in Tables~1 -- 4, we
included the ground-based optical points given in Price~et.al.~(2001),
as well as $J$ band data from \citet{dip+00}  and $J,H,$
$K^{\prime}$ and $K$-band points from \citet{fyn+01}. We
converted the $K^{\prime}$ points to $K$ using the prescription in
\citet{wc92}. All the data are corrected for absorption in our Galaxy,
and in the case of the optical data we subtracted the contribution
from the host.  The uncertainty in the radio flux is dominated by
interstellar scintillation over the timescales of interest here. In
calculating the $\chi^2$ for the fit, we add an uncertainty to the
radio fluxes based on the scintillation envelopes calculated using the
fireball size derived from the model.  We have excluded the first 8.46
data point from the fits, as it is likely associated with a separate
component from the reverse shock (see below).

We considered several possible cases in fitting the data; constant density
as well as a density gradient for the surrounding medium, and two cases for
the IC emission. The need to consider two cases arises from the fact that
for the same sychrotron emission spectrum, Sari \& Esin (2001) find two
possible IC solutions. These correspond to the two limits $
\eta\epsilon_e/\epsilon_B \equiv f << 1$, and $f >> 1$, where $\epsilon_e$
and $\epsilon_B$ are the fractions of the total explosion energy that go
into accelerating shocked electrons and amplifying the post-shock magnetic
field respectively. Here $\eta$ is the fraction of electron energy radiated
away, so describes whether cooling is dominant. If $f <<1$ (low IC), the IC
cooling rate is unimportant compared to that of synchrotron, whereas if $f
>> 1$ (high IC) inverse Compton cooling dominates the total emission.

Table~6 summarizes the physical parameters corresponding to the best
fit for each model.  All models we considered required the ejecta to
be collimated in order to explain the break in the optical lightcurve
at 1--2 days. In the optical, this break is roughly independent of
frequency.  Extinction of $A(V) \sim 0.1 - 0.2$ mag, presumably due to
the host galaxy, is required to fit the optical spectrum in three out
of four cases. We employed an LMC-like curve for the models, however
the magnitude of the extinction is relatively insensitive to the
extinction law. \citet{phg+01} derived similar values for collimation
and extinction based on fits to the optical data alone.

The model which best represents the broadband data is a constant
density ISM with the high-IC solution. Figures~\ref{fig1}a and
\ref{fig1}b show the broadband spectrum around day 2 and day 10, with
the best-fit model overplotted. The dot-dashed lines show the separate
contributions from the synchrotron and IC components.  Synchrotron
emission dominates at optical frequencies and below, however on both
day 2 and day 10 the model predicts an equal or larger contribution to
the X-ray data from IC scattering. From Figure~1 it is evident that
the approximation to the IC contribution made by assuming the same
shape as the synchrotron is not an issue in fitting the data. As
mentioned previously, the error in this approximation is only
significant below $\nu_a^{IC}$, and for the parameters we derive the
synchrotron dominates the total flux in this region by several orders
of magnitude.  Figures 2, 3, and 4 show the X-ray, optical and radio
lightcurves. The $ \chi^2$ for the fit is 124 for 114 degrees of
freedom.

The first column of data in Table~6 lists the physical parameters for
the best fit high-IC solution. The derived parameters are all
reasonable, although we note that $\epsilon_e$ of $0.30$ is only
marginally consistent with the assumption of adiabatic evolution
inherent to our model. While this will not alter our basic
conclusions, in future work we will expand the model to include
partially radiative blast waves (see
\citet{sar97} and
\citet{cps98} for treatments of the afterglow evolution in
this case).  

The isotropic-equivalent energy in the blast wave ($E_{iso}$ in
Table~6) of $1.8 \times 10^{53}$~erg is valid for $t \gsim 1$~day.  We
use an adiabatic model to fit the data, which assumes the total energy
is constant in time.  While this is approximately valid at $t \gsim 1
- 2$~days (i.e. for the majority of our data), the energy we derive
will not represent the total {\em initial} blastwave energy, since
significant energy is lost to radiation at early times.  In
particular, for the high-IC case, the shock wave radiates a
significant amount of energy early on, emitted as MeV
gamma-rays. Using the equations in Cohen, Piran \& Sari (1998) we
estimate the energy in the initial fireball to be about five times
higher than that associated with the adiabatically expanding blast
wave at a few days, or $E_{\rm initial} \sim 1 \times 10^{54}$~erg.
The observed isotropic energy release in the gamma-ray burst itself
was $E_{\gamma} = 3 \times 10^{53}$~erg
\citep{hmg+00}.  Recent
theoretical work suggests the internal shock process producing the GRB
should radiate less energy than the blast wave, except under extreme
physical conditions \citep{bel00,ks01}.  The value we derive for the
fireball energy is, therefore, reasonable.

To fit the low-IC ISM model, we initially allowed all of the physical
parameters to vary without limitation.  For the low-IC case, the
observed emission at all frequencies is from synchrotron radiation
alone. The fitting program adjusts the physical parameters in order to
reproduce all data with this component.  Reproducing the broadband
observations requires a large fraction of energy in the post-shock
magnetic field.  The best-fit value of $\epsilon_B = 24$ is
unphysical, and so we have fixed $\epsilon_B$ at unity -- the maximum
value physically possible.  For either value of $\epsilon_B$ the
low-IC model provides a poor fit to the data.  The $\chi^2$ is 187 for
114 datapoints for $\epsilon_B=1$ and 178 for 114 datapoints for
$\epsilon_B=24$.  The largest contribution to the $\chi^2$ comes from
the optical data, where the syncrotron component does not fit the data
as well as that derived for the high-IC case.  

\begin{deluxetable}{ccccc}
\footnotesize
\tablecolumns{5}
\tablewidth{0pc}
\tablecaption{Fit parameters for low and high-IC ISM and Wind models with 1--$\sigma$
errors.}\label{tab-model}
\tablehead{\colhead{Parameter} & \colhead{high-IC ISM} & \colhead{low-IC ISM} 
& \colhead{high-IC Wind} & \colhead{low-IC Wind}}
\startdata
$\chi^2$ for 114 data pts\tablenotemark{c} & 124 & 187 & 167 & 244  \\
 $t_{\rm jet}$ (days) & $1.55 \pm 0.14$ & $1.31 \pm 0.13$ & $2.53 \pm 0.45$ & $1.38 \pm 0.41$ \\
$t_{\rm non rel.}$ (days) & 70$ \pm 4$ & $112 \pm 7$ & $119 \pm 26 $ &
$308 \pm 152$ \\
$E_{\rm iso}\tablenotemark{a}$ ($10^{52}$ erg) & $18 \pm 2$ & $8.3 \pm 0.9$ 
& $43 \pm 7$ & $39 \pm 8$ \\
$n$(ISM)/$A_*$(wind)\tablenotemark{b} & 27$\pm 3$ cm$^{-3}$ & 1.6$\pm 0.2 $ cm$^{-3}$ &
3.5$\pm 0.4$ & 0.26$\pm 0.02$ \\
$p$  & 2.43$\pm .06$ & 2.20$\pm .04$ & 3.08$\pm .03$ & 2.25$\pm .03$ \\
$\epsilon_e$ (fraction of E) & 0.30$\pm .05$ & 0.16$\pm .02$ 
& 0.16$\pm .01$ & 0.018$\pm .003$ \\
$\epsilon_B$ (fraction of E) & 0.008$\pm .003$ & 1.0 & $ 0.005\pm .002$ & 1.0
\\
$\theta_{jet} (rad) $ & 0.137$\pm .004$ & 0.099$\pm .003$ 
& 0.103$\pm .002$ & .047$\pm .003$ \\
host A(V) & 0.12$\pm .02 $ & 0.20$\pm .02 $ & 0.00$\pm .01$ & 0.20$\pm .02$ \\
\enddata
\tablenotetext{a}{Isotropic equivalent blastwave energy (not corrected for collimation). See
text for details.}
\tablenotetext{b}{$\rho = A_*(5 \times 10^{11}) r^{-2}$g cm$^{-1}$}
\tablenotetext{c}{First 8.46 GHz excluded as an outlier.}
\end{deluxetable}

Table~6 provides the physical parameters associated with the best-fit
low-IC ISM model for $\epsilon_B$ fixed at unity.  In addition to the
large $\epsilon_B$, the isotropic-equivalent energy in the blast wave
of $10^{53}$~erg for the low-IC case is smaller than the observed
gamma-ray energy release.  Radiative corrections for this model are
much smaller than for the high-IC case, so that the adiabatic value
more closely represents the energy in the initial fireball.  As
pointed out above, $E_{\gamma} \gsim E_{\rm fireball}$ is hard to
accomodate except under extreme conditions.  

Figure~2 shows the X-ray data with both high and low-IC ISM models
overplotted.  The high-IC case provides a somewhat better fit to the
data, as the low-IC case systematically underpredicts the
measurements. The X-ray observations alone, however, are not
sufficient in this case to indicate the presence of an inverse Compton
component.  This requires the results from the broadband fit.  Our
calculations show that the IC
scattering hardens the X-ray spectrum over some time intervals
compared to the prediction low-IC model. There is only marginal
evidence for this in these data; the observed X-ray spectral photon
index is $1.9 \pm 0.21$ compared with the $p/2 +1 = 2.2$ associated
with the synchrotron component.  This emphasizes the role of future,
higher-significance X-ray observations in confirming the role if IC
scattering in GRB afterglows.

We also investigated a wind density profile in both high- and low-IC
regimes. Neither of these cases reproduce the data as well as the
high-IC ISM model. The low-IC wind solution has a poor $\chi^2$ of 244
for 114 degrees of freedom.  Again we have fixed $\epsilon_B=1.0$,
since the best-fit value of 174 is unphysical. The high-IC wind case
also has a large $\chi^2$ of 167 for 114 d.o.f, and under-predicts the
5~GHz radio data by a factor 2 -- 3.  It does, however, reproduce the
X-ray data reasonably well.  We note that early-time optical or
high-frequency radio data is particularly important in ruling out the
wind model. This is because the lightcurve evolution after the jet
break is similar for ISM and wind. Directly observing the cooling
break evolution provides the greatest leverage, since this evolution
is significantly different for wind and ISM models.  Generally, this
requires optical observations or high-frequency radio observations at
early epochs (minutes to hours after the event).  Unfortunately
optical coverage of this event began rather late ($t \sim 0.8$ days),
but future early-time observations of other events will provide a more
sensitive probe for density structure in the surrounding medium.

Finally, we point out that even for our best fit (high-IC ISM)
solution, the first 8 GHz data point exceeds the model's expectation
by a factor of 3.3 (7.9 $\sigma$). This discrepancy is worse for the
alternative models. Scintillation gains this large are extremely
improbable, and it is therefore likely that we are seeing evidence for
an additional component due to the reverse shock emission, as was
observed in early radio observations of GRB~990123 \citep{kfs+99}.

\section{Testing the High Density Model for GRB000926}

\begin{figure}
\plotone{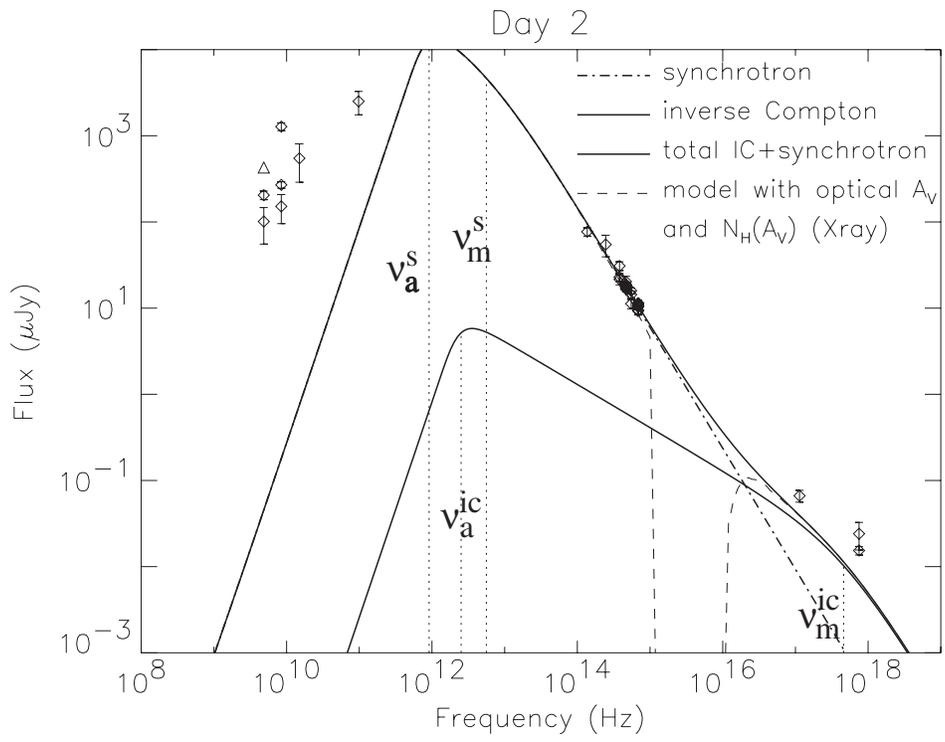}
\caption{Fit to the high-density ($n = 4 \times 10^4$~cm$^{-3}$) medium
of Piro {\em et al.} (2001).  The solid line shows the model calculations,
and the dashed lines show the synchrotron and inverse Compton contributions
separately.}
\end{figure}

Piro~{\em et al.} (2001) analyzed the X-ray and optical lightcurves of
GRB~000926, and noted that the spectral and temporal properties cannot
be explained using a standard jet model with moderate ISM density.  By
invoking inverse Compton cooling, we were able to fit the data with a
moderate density medium and relatively standard fireball parameters.
In order to describe the X-ray and optical lightcurves, Piro {\em et
al.} invoke moderate collimation of the fireball (opening angle
$\theta = 25^{\circ}$), and a dense ($n = 4 \times 10^4$~cm$^{-3}$)
medium.  This high density results in rapid deceleration of the
fireball, and an early transition to the non-relativistic regime.
These features 
seemingly reproduce the general characteristics of the high-frequency
observations. The high density found by Piro {\em et al.} would suggest
the event took place in a molecular cloud, and further the low extinction
measured in the optical would indicate that the GRB destroyed much of
the dust in its vicinity.

The radio data provide an important means of distinguishing these two
interpretations.  If the fireball expands in a very dense medium, the
ejecta decelerate much more rapidly, and at a given time the fireball
radius is significantly smaller than for the moderate densities
derived in this paper.  The self-absorption frequency, easily observed
as a cutoff in the radio spectrum, is correspondingly much higher at a
given time for the high-density case.

We have fit a model with the ISM density fixed at $n = 4 \times
10^4$~cm$^{-3}$, and the jet opening angle of $\theta = 25^{\circ}$ to
the broadband data set.  We investigated both high-IC and low-IC
constant-density solutions, and find we obtain the best fit using the
high-IC model.  Figure~5 shows the broadband spectrum on day 2, with a
comparison of our best fit with the data.  The radio data quite
clearly exclude the high-density possibility, since the
self-absorption frequency is more than a decade higher than we
measure.  On day two, the high density model predicts an 8~GHz radio
flux of $\sim0.2\mu$Jy, while we measure a flux of 200$\mu$Jy.  This
points out the importance of broadband coverage, including radio, in
constraining models for the GRB environment.

\section{Conclusions}

GRB~000926 has one of the highest-quality broadband afterglow
lightcurves studied to-date. We can explain all the general features
using a model of a relativistic shock produced by ejecta collimated to
a $\sim8^\circ$ opening angle. If we correct the isotropic-equivalent gamma-ray energy
release for this jet angle we get $E_{\gamma} = E_{\gamma , iso}
\times \theta^2/2 = 3 \times 10^{51}$~erg.  This is reasonably accounted
for in most currently-popular progenitor models. Our observations
require extinction in the host galaxy, with a corresponding hydrogen column of $\sim
2 \times 10^{20}$~cm$^{-2}$ if the dust to gas ratio is typical of the LMC
or SMC.  This is consistent with the event taking place in
a galactic disk.

We find that an explosion occurring in a spatially homogeneous medium
best describes the broadband lightcurves, provided that the ratio
$\eta \epsilon_e/\epsilon_B >> 1$. This condition implies that Compton
emission dominates the total cooling rate, and that IC scattering is
potentially directly observable in the afterglow spectrum. We do, in
fact, find that for our best-fit model IC emission dominates over the
synchrotron contribution at 1 keV on the 2 -- 10 day timescales
spanned by our X-ray data. Detecting the IC emission in the X-ray band
directly implies a lower-limit on the density \citep{pk00,se01}, in
this case $n \mathrel{\spose{\lower 3pt\hbox{$\mathchar"218$}} \raise
2.0pt\hbox{$\mathchar"13E$}} 10$ cm$^{-3}$, and our best fit yields $n
= 30$ cm$^{-3}$. This value for $n$ is higher than the average ISM
density in a typical galaxy, and is consistent with a diffuse
interstellar cloud, such as those commonly found in star forming
regions.  Combining the density with the host $N_H$ derived from our
optical observations (again assuming typical dust to gas ratios)
implies a scale size of 2~pc for the cloud.

\acknowledgements We wish to thank Harvey Tananbaum and the CXO operations
staff for facilitating the {\em Chandra} TOO observations. We thank Steve
Beckwith and the HST operations staff for facilitating the WFPC2
observations. FAH acknowledges support from a Presidential Early Career
award. SRK and SGD thank NSF for support of their ground-based GRB
programs. RS is grateful for support from a NASA ATP grant.
RS \& TJG acknowledge support from the Sherman Fairchild Foundation.


\end{document}